\documentclass[12pt]{iopart}
\usepackage{amssymb,accents}

\pdfoutput=1

\newcommand{\wt}{\tilde}
\newcommand{\wh}{\hat}
\newcommand{\wb}{\bar}
\newcommand{\ut}[1]{{\underaccent{\tilde}{#1}}}
\newcommand{\uh}[1]{{\underaccent{\hat}{#1}}}
\renewcommand{\th}[1]{\wh{\wt{#1}}}
\newcommand{\hb}[1]{\wb{\wh{#1}}}
\newcommand{\bt}[1]{\wt{\wb{#1}}}
\newcommand{\thb}[1]{\wh{\wt{\wb{#1}}}}

\newcommand{\sn}{\mathop{\mathrm{sn}}\nolimits}

\newcommand{\cQ}{\mathcal{Q}}
\newcommand{\cF}{\mathcal{F}}
\newcommand{\oT}{\mathsf{T}}
\newcommand{\cB}{\mathcal{B}}

\newcommand{\textonehalf}{$\textstyle \frac{1}{2}$}

\newtheorem{proposition}{Proposition}

\begin{document}
\paper[Multi-quadratic quad equations: integrable cases]{Multi-quadratic quad equations: integrable cases from a factorised-discriminant hypothesis}
\author{James Atkinson}
\address{School of Mathematics and Statistics, The University of Sydney, NSW 2006, Australia.}
\author{Maciej Nieszporski}
\address{Katedra Metod Matematycznych Fizyki, Wydzial Fizyki, Uniwersytet Warszawski, ul. Ho\.za 74, Poland.}
\date{\today}
\begin{abstract}
We give integrable quad equations which are multi-quadratic (degree-two) counterparts of the well-known multi-affine (degree-one) equations classified by Adler, Bobenko and Suris (ABS).
These multi-quadratic equations define multi-valued evolution from initial data, but our construction is based on the hypothesis that discriminants of the defining polynomial factorise in a particular way that allows to reformulate the equation as a single-valued system.
Such reformulation comes at the cost of introducing auxiliary (edge) variables and augmenting the initial data.
Like the multi-affine equations listed by ABS, these new models are consistent in multidimensions.
We clarify their relationship with the ABS list by obtaining B\"acklund transformations connecting all but the primary multi-quadratic model back to equations from the multi-affine class.
\end{abstract}
\section{Introduction}
We consider quad equations defined in terms of a polynomial in four variables, $\cQ$, that is equations of the form
\begin{equation}
\cQ(u,\wt{u},\wh{u},\th{u})=0,\label{ge}
\end{equation}
where in the simplest setting $u=u(n,m)$, $\wt{u}=u(n+1,m)$, $\wh{u}=u(n,m+1)$ and $\th{u}=u(n+1,m+1)$ are values of a dependent variable taking values in $\mathbb{C}\cup\{\infty\}$ as a function of independent variables $n,m\in\mathbb{Z}$.
The quad equation is called multi-affine if the defining polynomial, $\cQ$, is degree one in each variable, and we call it multi-quadratic if $\cQ$ is degree two in each variable.

An important integrability feature that is possible for the quad equation is the multidimensional consistency \cite{FrankABS,BS}; it has proven to be a natural property of many integrable equations in the multi-affine class.
Significant work on this is the list of multi-affine quad equations with the consistency property obtained by Adler, Bobenko and Suris (ABS) in \cite{ABS,ABS2}.
Quad equations beyond the multi-affine class were considered in \cite{KaNie} where several multidimensionally consistent examples (also beyond the multi-quadratic) were obtained. 
In fact these equations appeared naturally in relation to underlying models in a different class, namely the Yang-Baxter maps \cite{Ves,drin}, and emerge from a generalisation of results connecting the Yang-Baxter maps with the multi-affine quad equations \cite{ABS,PSTV,Tasos}. 
One multidimensionally consistent multi-quadratic quad equation identified in \cite{KaNie} was known earlier due to observations of Adler and Veselov in \cite{AdVeQ}, it in fact arises as the superposition principle for B\"acklund transformations of the KdV equation. 
Recently it has also come to light that the well-known discrete version of the KdV equation due to Hirota \cite{hirota-0}, which is multi-affine but absent from the ABS list, is naturally understood within the consistency framework as a special case of a multi-quadratic quad equation \cite{JamesQ}.
The few known examples therefore indicate that multi-quadratic quad equations are a quite natural and potentially rich class of integrable systems.

A feature of higher degree discrete models, like the multi-quadratic quad equations, is that they define multivalued evolution from initial data. 
This adds richness, but also another level of difficulty in dealing with such systems.
This difficulty is however mitigated in the mentioned integrable cases because they can be reformulated as a single-valued system with augmented initial data.
For the models in \cite{KaNie} this is because the variables present in the associated Yang-Baxter map themselves satisfy a single-valued system.
But that situation can actually be viewed as a special case of a more general constructive approach to this kind of reformulation \cite{JamesQ}.
In this approach auxiliary variables are introduced on lattice edges, which are similar to variables of an associated Yang-Baxter map, however, rather than satisfying an independent system, they instead participate in a mixed system involving both vertex and edge variables.
The resulting model is of a type similar to the class introduced in \cite{hv11} but with the additional feature of preserving algebraic relations on the lattice edges.
The reformulation procedure relies completely on a discriminant factorisation property of the defining polynomial.
It is this special property that is the departure point and the main emphasis of the present work, we use it as a foundation to enable some systematic investigations of the multi-quadratic quad equations.

The main result presented here is a list of multi-quadratic quad equations with the discriminant factorisation property.
To construct the models we start from biquadratic polynomials (the factors of the discriminant) which are associated with the edges of the lattice, and this provides a natural correspondence with the multi-affine ABS equations whose construction also involves edge biquadratics.
Due to the factorised discriminant hypothesis the models we list admit reformulation as single-valued systems.
However the more important property of these equations, which is actually rather remarkable because it is not explicitly built into the construction, is the integrability feature of multidimensional consistency.

The second part of our paper is devoted to developing the transformation theory of the obtained models.
The most pressing question we seek to address relates to existence of transformations of B\"acklund or Miura type connecting these multi-quadratic models back to the better studied multi-affine ABS equations.
Such transformations are not inherent from our construction, rather we tackle the problem of obtaining transformations a posteriori using separate methods.
The main result is to connect all except the primary model, namely the multi-quadratic counterpart of Q4, back to equations in the multi-affine (ABS) class.

We proceed as follows.
In Section \ref{dfh} we explain the basic discriminant factorisation hypothesis for the multi-quadratic quad equations.
Additional assumptions are explained in Section \ref{assumptions} and the obtained models are listed in Section \ref{list}.
The method to reformulate these models as systems that define single-valued evolution from initial data is given in Section \ref{SVS} and their multidimensional consistency is described in Section \ref{MDC}.
Section \ref{transformations} is devoted to the transformation theory of the models, in particular we recall the multi-affine equations from the ABS list in Section \ref{ABSlist} and give transformations connecting the multi-quadratic models to them in Section \ref{BTex}.
The methods used to obtain these transformations are explained in Sections \ref{nsd}, \ref{idolons} and \ref{quadlin}.
Some questions raised by the results reported here are discussed in Section \ref{discussion}.

\section{The factorised-discriminant hypothesis}\label{dfh}
In the multi-quadratic case considered here, the defining polynomial $\cQ$ of the quad equation (\ref{ge}) is of degree two in each of the four variables.
To calculate the evolution defined by (\ref{ge}) from some initial data requires solving this equation locally as a quadratic equation for one of the arguments, and in particular taking the square root of its discriminant with respect to that argument.
In general this discriminant is a polynomial of degree four in each of the remaining three arguments; the property of polynomial $\cQ$ that we study here is the factorisation of this discriminant into a product of three factors:
\begin{equation}
\eqalign{
\Delta[\cQ(u,\wt{u},\wh{u},\th{u}),\th{u}] = H_1(u,\wt{u})H_2(u,\wh{u})G_1(\wt{u},\wh{u}),\\
\Delta[\cQ(u,\wt{u},\wh{u},\th{u}),\wh{u}] = H_1(u,\wt{u})H_2(\wt{u},\th{u})G_2(u,\th{u}),\\
\Delta[\cQ(u,\wt{u},\wh{u},\th{u}),\wt{u}] = H_1(\wh{u},\th{u})H_2(u,\wh{u})G_3(u,\th{u}),\\
\Delta[\cQ(u,\wt{u},\wh{u},\th{u}),    u ] = H_1(\wh{u},\th{u})H_2(\wt{u},\th{u})G_4(\wt{u},\wh{u}),
}\label{df}
\end{equation}
where $H_1$ and $H_2$ are degree-two polynomials in each of two variables, i.e., biquadratics, and $G_1,\ldots,G_4$ are degenerate biquadratics of the form $G_i(\wt{u},\wh{u})=(a_i+b_i\wt{u}+c_i\wh{u}+d_i\wt{u}\wh{u})^2$, $i\in\{1,2,3,4\}$, i.e., the square of a polynomial which is degree-one in each variable.
This is the most general hypothesis that allows replacement of (\ref{ge}) by an associated single-valued system through introduction of auxiliary edge variables $\sigma_1$ and $\sigma_2$ satisfying the edge relations
\begin{equation}
\sigma_1^2 = H_1(u,\wt{u}), \quad \sigma_2^2 = H_2(u,\wh{u}).\label{er}
\end{equation}
The key features allowing this are that all discriminants in (\ref{df}) are squares which clearly leads locally to a rational model, and furthermore that the edge relations on opposite sides of the quad are the same, allowing the rational model to replace the quad equation globally (for instance in the simplest setting throughout $\mathbb{Z}^2$).
The procedure to obtain the single-valued system is straightforward, it was explained in detail in \cite{JamesQ} and an example will also be included later in this paper (Section \ref{SVS}).

We remark that the restriction to $G_1,\ldots, G_4$ being degenerate seems unnatural in some regards, in fact there exist sets of polynomials $\{\cQ,H_1,H_2,H_3\}$ satisfying system (\ref{df}) with $G_1=G_2=G_3=G_4=H_3$, and where $H_3$ is also non-degenerate.
These more symmetric solutions of the problem are interesting, however to understand the sense in which the resulting polynomial $\cQ$ defines an integrable discrete model requires a substantial alteration of the setting, and this will be studied in detail elsewhere.
Also we remark that the assumed symmetry can be relaxed, for instance replacing $H_1$ appearing in the third and fourth equations of (\ref{df}) with $\wh{H}_1\neq H_1$.
Such systems are of interest too but the associated polynomial $\cQ$ in that case is more naturally interpreted as defining a B\"acklund transformation between models.

\section{Additional assumptions}\label{assumptions}
The solution of (\ref{df}), in the sense of obtaining a set of polynomials \[\{\cQ,H_1,H_2,G_1,G_2,G_3,G_4\}\] for which system (\ref{df}) is identically satisfied, is a difficult classification problem in its full generality.
Further assumptions allow this problem to be solved using computer algebra, these come from looking at examples identified in \cite{JamesQ}.

We assume invariance of $\cQ$ when the arguments are permuted by the Klein four-group (the Kleinian symmetry)
\begin{equation}
\cQ(u,\wt{u},\wh{u},\th{u})=\cQ(\wt{u},u,\th{u},\wh{u})=\cQ(\wh{u},\th{u},u,\wt{u}).
\end{equation}
This reduces the generic multi-quadratic polynomial $\cQ$ to the form
\begin{equation}
\fl
\eqalign{
\cQ(u,\wt{u},\wh{u},\th{u})=
a_{{1}}  +a_{{2}} \left( u+\wt{u}+\wh{u}+\th{u} \right)+a_{{3}} \left( {u}^{2}+{\wt{u}}^{2}+{\wh{u}}^{2}+{\th{u}}^{2} \right) 
+a_{{4}} \left( \wh{u}\th{u}+u\wt{u} \right) 
\\\quad
+a_{{5}} \left( \wt{u}\wh{u}+u\th{u} \right) +a_{{6}} \left( \wt{u}\th{u}+u\wh{u} \right)
+a_{{7}} \left( \wt{u}\wh{u}\th{u}+u\wh{u}\th{u}+u\wt{u}\th{u}+u\wt{u}\wh{u} \right) 
\\\quad
+a_{{8}} \left( u{\wt{u}}^{2}+\wh{u}{\th{u}}^{2}+{u}^{2}\wt{u}+{\wh{u}}^{2}\th{u} \right) 
+a_{{9}} \left( u{\wh{u}}^{2}+{u}^{2}\wh{u}+\wt{u}{\th{u}}^{2}+{\wt{u}}^{2}\th{u} \right)
\\\quad
+a_{{10}} \left( u{\th{u}}^{2}+{u}^{2}\th{u}+\wt{u}{\wh{u}}^{2}+{\wt{u}}^{2}\wh{u} \right) 
+a_{{11}}u\wt{u}\wh{u}\th{u}
+a_{{12}} \left( \wt{u}\th{u}+u\wh{u} \right)  \left( \wt{u}\wh{u}+u\th{u} \right) 
\\\quad
+a_{{13}} \left( \wt{u}\wh{u}+u\th{u} \right)  \left( \wh{u}\th{u}+u\wt{u} \right)
+a_{{14}} \left( \wt{u}\th{u}+u\wh{u} \right)  \left( \wh{u}\th{u}+u\wt{u} \right)
+a_{{15}} \left( {\wh{u}}^{2}{\th{u}}^{2}+{u}^{2}{\wt{u}}^{2} \right) 
\\\quad
+a_{{16}} \left( {\wt{u}}^{2}{\wh{u}}^{2}+{u}^{2}{\th{u}}^{2} \right) 
+a_{{17}} \left( {\wt{u}}^{2}{\th{u}}^{2}+{u}^{2}{\wh{u}}^{2} \right)
+a_{{18}}u\wt{u}\wh{u}\th{u} \left( u+\wt{u}+\wh{u}+\th{u} \right) 
\\\quad
+a_{{19}} \left( \wt{u}{\wh{u}}^{2}{\th{u}}^{2}+{u}^{2}{\wt{u}}^{2}\th{u}+u{\wh{u}}^{2}{\th{u}}^{2}+{u}^{2}{\wt{u}}^{2}\wh{u} \right) 
+a_{{20}} \left( {\wt{u}}^{2}\wh{u}{\th{u}}^{2}+u{\wt{u}}^{2}{\th{u}}^{2}+{u}^{2}{\wh{u}}^{2}\th{u}+{u}^{2}\wt{u}{\wh{u}}^{2} \right) 
\\\quad
+a_{{21}} \left( {\wt{u}}^{2}{\wh{u}}^{2}\th{u}+u{\wt{u}}^{2}{\wh{u}}^{2}+{u}^{2}\wh{u}{\th{u}}^{2}+{u}^{2}\wt{u}{\th{u}}^{2} \right)
+a_{{22}}u\wt{u}\wh{u}\th{u} \left( \wh{u}\th{u}+u\wt{u} \right) 
\\\quad
+a_{{23}}u\wt{u}\wh{u}\th{u} \left( \wt{u}\wh{u}+u\th{u} \right) 
+a_{{24}}u\wt{u}\wh{u}\th{u} \left( \wt{u}\th{u}+u\wh{u} \right) 
\\\quad
+a_{{25}} \left( {\wt{u}}^{2}{\wh{u}}^{2}{\th{u}}^{2}+{u}^{2}{\wh{u}}^{2}{\th{u}}^{2}+{u}^{2}{\wt{u}}^{2}{\th{u}}^{2}+{u}^{2}{\wt{u}}^{2}{\wh{u}}^{2} \right) 
\\\quad
+a_{{26}}u\wt{u}\wh{u}\th{u} \left( \wt{u}\wh{u}\th{u}+u\wh{u}\th{u}+u\wt{u}\th{u}+u\wt{u}\wh{u} \right) 
+a_{{27}}{u}^{2}{\wt{u}}^{2}{\wh{u}}^{2}{\th{u}}^{2},
}
\end{equation}
for some set of coefficients $\{a_1,\ldots,a_{27}\}$.
This symmetry is a strong additional assumption, for instance by itself it is actually sufficient for integrability of quad equations in the multi-affine class \cite{ABS2,via}.

We also assume that both biquadratic polynomials $H_1$ and $H_2$ are symmetric, and furthermore that they are taken from one of the following one-parameter families (the parameter being denoted by $p$)
\begin{equation}
\label{q4}
\fl\textrm{q4}^*:\quad\frac{c}{2}\left(1+u^2\wt{u}^2+\wt{u}^2p^2+p^2u^2\right)-\frac{1}{2c}\left(u^2\wt{u}^2p^2+u^2+\wt{u}^2+p^2\right)-\left(c^2-\frac{1}{c^2}\right)u\wt{u}p,
\end{equation}
\begin{equation}
\label{q3}
\fl\textrm{q3}^*:\quad\frac{1}{2}(u^2+\wt{u}^2)+\frac{\delta^2}{2}(p^2-1)-u\wt{u}p,
\end{equation}
\begin{equation}
\label{q2}
\fl\textrm{q2}^*:\quad\frac{1}{4}\left(u^2+\wt{u}^2+p^2\right)-\frac{1}{2}\left(u\wt{u}+\wt{u}p+pu\right),
\end{equation}
\begin{equation}
\label{q1}
\fl\textrm{q1}^*:\quad (\wt{u}-u)^2-p,
\end{equation}
\begin{equation}
\label{a2}
\fl\textrm{a2}^*:\quad\frac{1}{2}(1+u^2\wt{u}^2)-pu\wt{u},
\end{equation}
\begin{equation}
\fl\textrm{a1}^*:\quad\label{a1}(\wt{u}+u)^2-p,
\end{equation}
\begin{equation}
\label{h3}
\fl\textrm{h3}^*:\quad u\wt{u}p+\delta^2,
\end{equation}
\begin{equation}
\label{h2}
\fl\textrm{h2}^*:\quad u+\wt{u}+p.
\end{equation}
Where it appears $\delta \in\{0,1\}$, whilst in (\ref{q4}) $c\in\mathbb{C}\setminus\{0,1,-1,i,-i\}$ is a fixed constant.
Up to a point transformation of the parameters these biquadratic polynomials coincide with those associated with the ABS equations \cite{ABS,ABS2} (precise parameter associations will be given later in Section \ref{ABSlist}).
Note that (\ref{h3}) and (\ref{h2}) are considered to be biquadratics here also, but this is in a projective sense, so loosely speaking we view for instance (\ref{h2}) as the polynomial $(u+\wt{u}+p)(u-\infty)(\wt{u}-\infty)$.
We remark that by a M\"obius change of variables a symmetric biquadratic polynomial can always be brought to one of the forms (\ref{q4})--(\ref{h2}) for some choice of parameter $p$, or else is the square of a multi-affine polynomial (like $G_1,\ldots,G_4$ above), a possibility which we therefore exclude by restricting to (\ref{q4})--(\ref{h2}).
The discriminant of biquadratics (\ref{q4})--(\ref{h2}) with respect to $\wt{u}$ is a polynomial of degree at most four in $u$, the main characterising feature of the biquadratic families is that roots of this discriminant polynomial do not change upon altering parameter $p$.

Finally, the assumed symmetry of $\cQ$ together with (\ref{df}) implies already that \[G_1=G_2=G_3=G_4=:G,\] in fact we will go further and make the assumption that 
\begin{equation}
G(\wt{u},\wh{u}) = (\wt{u}-\wh{u})^2,
\end{equation}
which is our last additional assumption.

\section{The obtained multi-quadratic quad equations}\label{list}
The additional assumptions enable solution of the discriminant factorisation hypothesis directly by computer algebra, which gives the main result of our paper as follows.
\begin{proposition}\label{listprop}
Let $\cQ$ be a polynomial of degree two in each of four variables with the Kleinian symmetry 
\begin{equation}
\cQ(u,\wt{u},\wh{u},\th{u})=\cQ(\wt{u},u,\th{u},\wh{u})=\cQ(\wh{u},\th{u},u,\wt{u}),
\end{equation}
such that the discriminants factorise as follows
\begin{equation}
\Delta[\cQ(u,\wt{u},\wh{u},\th{u}),\th{u}]\propto (\wt{u}-\wh{u})^2H_1(u,\wt{u})H_2(u,\wh{u}),\label{disc}
\end{equation}
where $H_1$ and $H_2$ are biquadratic polynomials.
If $H_1$ and $H_2$ are taken from one of the biquadratic families (\ref{q4})--(\ref{h2}) with generic parameter choices denoted $p$ and $q$ respectively, then $\cQ$ is determined up to an overall constant and defines respectively the quad equations (\ref{QQ4})--(\ref{QH2}) below.
\end{proposition}
Q4$^*$:
\begin{equation}
\eqalign{
\fl\quad (p-q)[(c^{-2}p-c^2q)(u\wt{u}-\wh{u}\th{u})^2-(c^{-2}q-c^2p)(u\wh{u}-\wt{u}\th{u})^2]\\
-(p-q)^2[(u+\th{u})^2(1+\wt{u}^2\wh{u}^2)+(\wt{u}+\wh{u})^2(1+u^2\th{u}^2)]\\
+[(u-\th{u})(\wt{u}-\wh{u})(c^{-1}-cpq)-2(p-q)(1+u\wt{u}\wh{u}\th{u})]\\
\times[(u-\th{u})(\wt{u}-\wh{u})(c^{-1}pq-c)-2(p-q)(u\th{u}+\wt{u}\wh{u})]=0,\\
}\label{QQ4}
\end{equation}
Q3$^*$:
\begin{equation}
\eqalign{
\fl\quad (p-q)[p(u\wt{u}-\wh{u}\th{u})^2-q(u\wh{u}-\wt{u}\th{u})^2]-\delta^2(p-q)^2[(u+\th{u})^2+(\wt{u}+\wh{u})^2]\\
\fl\quad +[(u-\th{u})(\wt{u}-\wh{u})-2\delta^2(p-q)][(u-\th{u})(\wt{u}-\wh{u})(pq-1)-2(p-q)(u\th{u}+\wt{u}\wh{u})]=0,
}\label{QQ3}
\end{equation}
Q2$^*$:
\begin{equation}
\eqalign{
\fl\quad (p-q)[p(u\wt{u}-\wh{u}\th{u})(u+\wt{u}-\wh{u}-\th{u})-q(u\wh{u}-\wt{u}\th{u})(u+\wh{u}-\wt{u}-\th{u})] \\
+(u-\th{u})(\wt{u}-\wh{u})[p(u-\wh{u})(\wt{u}-\th{u})-q(u-\wt{u})(\wh{u}-\th{u})-pq(p-q)]=0,
}\label{QQ2}
\end{equation}
Q1$^*$:
\begin{equation}
\eqalign{
\fl\quad (p-q)[p(u+\wt{u}-\wh{u}-\th{u})^2-q(u+\wh{u}-\wt{u}-\th{u})^2]\\+4(u-\th{u})(\wt{u}-\wh{u})[p(u-\wh{u})(\wt{u}-\th{u})-q(u-\wt{u})(\wh{u}-\th{u})]=0,
}\label{QQ1}
\end{equation}
A2$^*$:
\begin{equation}
\eqalign{
\fl\quad (p-q)[p(u\wh{u}-\wt{u}\th{u})^2-q(u\wt{u}-\wh{u}\th{u})^2]\\
+(u-\th{u})(\wt{u}-\wh{u})[(u-\th{u})(\wt{u}-\wh{u})(pq-1)+2(p-q)(1+u\wt{u}\wh{u}\th{u})]=0,
}\label{QA2}
\end{equation}
A1$^*$:
\begin{equation}
\eqalign{
\fl\quad (p-q)[p(u-\wt{u}+\wh{u}-\th{u})^2-q(u-\wh{u}+\wt{u}-\th{u})^2]\\+4(u-\th{u})(\wt{u}-\wh{u})[p(u+\wh{u})(\wt{u}+\th{u})-q(u+\wt{u})(\wh{u}+\th{u})]=0,
}\label{QA1}
\end{equation}
H3$^*$:
\begin{equation}
\eqalign{
\fl\quad (p-q)[p(u\wh{u}-\wt{u}\th{u})^2-q(u\wt{u}-\wh{u}\th{u})^2]\\
+(u-\th{u})(\wt{u}-\wh{u})[(u-\th{u})(\wt{u}-\wh{u})pq-4\delta^2(p-q)]=0,
}\label{QH3}
\end{equation}
H2$^*$:
\begin{equation}
\eqalign{
\fl\quad (p-q)[p(u-\wt{u}+\wh{u}-\th{u})^2-q(u-\wh{u}+\wt{u}-\th{u})^2]\\+(u-\th{u})(\wt{u}-\wh{u})[(u-\th{u})(\wt{u}-\wh{u})-2(p-q)(u+\wt{u}+\wh{u}+\th{u})]=0.
}\label{QH2}
\end{equation}

We have denoted the list of models here $Q4^*,\ldots,H2^*$ due to their natural correspondence with equations from the ABS list, which are denoted $Q4,\ldots,H1$ \cite{ABS}. (The ABS list and its precise relationship with (\ref{QQ4})--(\ref{QH2}) will be given in Section \ref{ABSlist}.)
The biquadratic polynomials associated with ABS equations $Q1_{\delta=0}$, $A1_{\delta=0}$ and $H1$ are already squares, which means the multi-quadratic counterparts of those models factorise into products of multi-affine polynomials, cases we exclude here as degenerate.

Similar to the situation for ABS equations, all of (\ref{QQ3})--(\ref{QH2}) can be obtained from the primary model (\ref{QQ4}) by limiting procedures. 
All listed models are non-equivalent by autonomous M\"obius changes of variables and point transformations of the parameters.
However, the models $A2^*$ and $Q3^*_{\delta=0}$ are related by the non-autonomous point transformation $u\rightarrow u^{(-1)^{n+m}}$, whilst $A1^*$ and $Q1^*$ are related by $u\rightarrow (-1)^{n+m}u$.

To our knowledge the equations in the list (\ref{QQ4})--(\ref{QH2}) are new except for the following cases.
The model H2$^*$ (\ref{QH2}) appeared as the superposition principle for solutions of the KdV equation in \cite{AdVeQ} and in relation to Yang-Baxter maps in \cite{KaNie}.
The model A1$^*$ (\ref{QA1}) was obtained, although not in explicit form, in \cite{KaNie3}.
The model A2$^*$ (\ref{QA2}) was obtained originally in \cite{JamesQ} as the superposition principle for B\"acklund transformations of Hirota's discrete KdV equation \cite{hirota-0}.

\section{Reformulation as single-valued systems}\label{SVS}
One of the most salient features of quad equations from the multi-quadratic class is that they define multivalued evolution from initial data.
However, built into the construction of models (\ref{QQ4})--(\ref{QH2}) is the discriminant factorisation property, and this allows the quad equation to be reformulated as a system that defines single-valued evolution from augmented initial data \cite{JamesQ}.
Here we perform this reformulation on the primary model Q4$^*$ (\ref{QQ4}), and describe the relationship between the quad equation and its reformulation.

\begin{figure}[t]
\begin{center}
\begin{picture}(100,100)(0,0)
\multiput(21,12)(60,0){2}{{\line(0,1){60}}}
\multiput(21,12)(0,60){2}{{\line(1,0){60}}}
\put(7,39){{$\sigma_2$}}
\put(48,0){{$\sigma_1$}}
\put(87,39){{$\wt{\sigma}_2$}}
\put(48,78){{$\wh{\sigma}_1$}}
\put(21,12){\circle*{3}}
\put(81,12){\circle*{3}}
\put(21,72){\circle*{3}}
\put(81,72){\circle*{3}}
\put(12,4){$u$}
\put(84,4){$\wt{u}$}
\put(12,72){$\wh{u}$}
\put(84,72){$\th{u}$}
\end{picture}
\end{center}
\caption{Variables assigned to the vertices of a quadrilateral, and auxiliary variables to the edges.}
\label{quadpic}
\end{figure}
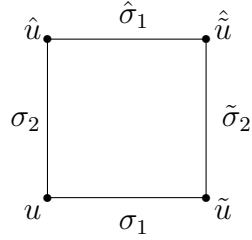

As described in Section \ref{dfh}, the auxiliary variables on lattice edges (see Figure \ref{quadpic}) are introduced through the relations
\begin{equation}
\eqalign{
\fl \sigma_1^2 = \frac{c}{2}\left(1+u^2\wt{u}^2+\wt{u}^2p^2+p^2u^2\right)-\frac{1}{2c}\left(u^2\wt{u}^2p^2+u^2+\wt{u}^2+p^2\right)-\left(c^2-\frac{1}{c^2}\right)u\wt{u}p,\\
\fl \sigma_2^2 = \frac{c}{2}\left(1+u^2\wh{u}^2+\wh{u}^2q^2+q^2u^2\right)-\frac{1}{2c}\left(u^2\wh{u}^2q^2+u^2+\wh{u}^2+q^2\right)-\left(c^2-\frac{1}{c^2}\right)u\wh{u}q,
}\label{dvdef}
\end{equation}
which are in terms of the associated biquadratic polynomial given earlier in (\ref{q4}).
By solving equation (\ref{QQ4}) as a quadratic equation for $\th{u}$ and exploiting these edge variables we obtain an equation of the form
\begin{equation}
\cF(u,\wt{u},\wh{u},\th{u},\sigma_1\sigma_2)=0
\end{equation}
which is by construction of polynomial degree one in $\th{u}$ and $\sigma_1\sigma_2$, for instance:
\begin{equation}
\eqalign{
\fl\quad\cF(u,\wt{u},\wh{u},\th{u},\sigma_1\sigma_2):=u[(c^{-2}p-c^2q)(u\wt{u}-\wh{u}\th{u})+(c^{-2}q-c^2p)(u\wh{u}-\wt{u}\th{u})]\\
-(u-\th{u})[(c^{-1}-cpq)(u^2+\wt{u}\wh{u})+(c^{-1}pq-c)(1+u^2\wt{u}\wh{u})+2\sigma_1\sigma_2]\\
-(p-q)(\wt{u}-\wh{u})(1+u^3\th{u})
}\label{Fdef}
\end{equation}
(the precise form of this expression is chosen because it is simple, but it is not unique).
Sequentially solving for each of the quad variables of (\ref{QQ4}) in the same way leads to the following system
\begin{equation}
\eqalign{
\cF(u,\wt{u},\wh{u},\th{u},\sigma_1\sigma_2)=0,\\
\cF(\wt{u},u,\th{u},\wh{u},\sigma_1\wt{\sigma}_2)=0,\\
\cF(\wh{u},\th{u},u,\wt{u},\wh{\sigma}_1\sigma_2)=0,\\
\cF(\th{u},\wh{u},\wt{u},u,\wh{\sigma}_1\wt{\sigma}_2)=0,
}
\label{quadsys}
\end{equation}
up to the choice of sign of the discriminant terms appearing in the last argument.

The system (\ref{quadsys}), (\ref{dvdef}) is the single-valued model associated with (\ref{QQ4}). 
It is easily verifiable that each equation in (\ref{quadsys}) implies (\ref{QQ4}) is satisfied modulo the relations (\ref{dvdef}).
The signs of the discriminant terms in (\ref{quadsys}) have been chosen for self-consistency (ensuring that any one equation in (\ref{quadsys}) is a consequence of the other three), to preserve the Kleinian symmetry, and to preserve the consistency property described in the following section.

The usual initial value problem for a quad-equation involves specifying the dependent variable on vertices along some admissible lattice path or collection of paths.
The modification required for system (\ref{quadsys}) is that variables on path edges, subject to the constraint (\ref{dvdef}), also need to be specified.
The system (\ref{quadsys}), (\ref{dvdef}) is therefore actually a model of vertex-bond type as described in \cite{hv11}, but it has the additional feature of preserving algebraic relations (\ref{dvdef}) on lattice edges.
In this setting the multivalued evolution defined by equation (\ref{QQ4}) is reflected in the non-uniqueness of initial edge variables when they are defined in terms of the initial vertex variables through (\ref{dvdef}).

In Table \ref{svms} the polynomial $\cF$ that appears in the single-valued system associated with each equation from the list (\ref{QQ4})--(\ref{QH2}) is also given.
The auxiliary variables present in the table are introduced through the edge relations which can be written generically as
\begin{equation}
\sigma_1^2 = H_1(u,\wt{u}), \quad \sigma_2^2 = H_2(u,\wh{u}),\label{erer}
\end{equation}
in terms of the associated biquadratic polynomials (cf. Proposition \ref{listprop}).

\begin{table}
\begin{tabular}{rl}
\hline
Q4$^*$
& $u[(c^{-2}p-c^2q)(u\wt{u}-\wh{u}\th{u})+(c^{-2}q-c^2p)(u\wh{u}-\wt{u}\th{u})]$\\
& \quad $-(u-\th{u})[(c^{-1}-cpq)(u^2+\wt{u}\wh{u})+(c^{-1}pq-c)(1+u^2\wt{u}\wh{u})+2\sigma_1\sigma_2]$\\
& \quad $-(p-q)(\wt{u}-\wh{u})(1+u^3\th{u})$\\
Q3$^*$
& $u[p(u\wt{u}-\wh{u}\th{u})+q(u\wh{u}-\wt{u}\th{u})]-(u-\th{u})[u^2+\wt{u}\wh{u}-\delta^2(1-pq)-2\sigma_1\sigma_2]$\\
& \quad $-\delta^2(p-q)(\wt{u}-\wh{u})$ \\
Q2$^*$
& $u[p(u+\wt{u}-\wh{u}-\th{u})+q(u-\wt{u}+\wh{u}-\th{u})]+p(u\wt{u}-\wh{u}\th{u})+q(u\wh{u}-\wt{u}\th{u})$\\
& \quad $-(u-\th{u})[(u-\wt{u})(u-\wh{u})+pq+4\sigma_1\sigma_2]$\\
Q1$^*$
& $p(u+\wt{u}-\wh{u}-\th{u})+q(u-\wt{u}+\wh{u}-\th{u})-2(u-\th{u})[(u-\wt{u})(u-\wh{u})-\sigma_1\sigma_2]$\\
A2$^*$
& $u[p(u\wh{u}-\wt{u}\th{u})+q(u\wt{u}-\wh{u}\th{u})]-(u-\th{u})(1+u^2\wt{u}\wh{u}-2\sigma_1\sigma_2)$\\
A1$^*$
& $p(u+\wh{u}-\wt{u}-\th{u})+q(u-\wh{u}+\wt{u}-\th{u})-2(u-\th{u})[(u+\wt{u})(u+\wh{u})-\sigma_1\sigma_2]$\\
H3$^*$
& $u[p(u\wh{u}-\wt{u}\th{u})+q(u\wt{u}-\wh{u}\th{u})]+2(u-\th{u})(\delta-\sigma_1\sigma_2)$\\
H2$^*$
& $p(u+\wh{u}-\wt{u}-\th{u})+q(u-\wh{u}+\wt{u}-\th{u})+(u-\th{u})(\wt{u}+\wh{u}+2u-2\sigma_1\sigma_2)$\\
\hline
\end{tabular}
\caption{
Polynomials $\cF(u,\wt{u},\wh{u},\th{u},\sigma_1\sigma_2)$ that define the single-valued system (\ref{quadsys}) associated with the multi-quadratic models (\ref{QQ4})--(\ref{QH2}) of Proposition \ref{listprop}.
The auxiliary variables $\sigma_1$ and $\sigma_2$ satisfy edge relations $\sigma_1^2=H_1(u,\wt{u})$ and $\sigma_2^2=H_2(u,\wh{u})$ in terms of the associated biquadratic polynomials listed in (\ref{q4})--(\ref{h2}).
}\label{svms}
\end{table}

The single-valued reformulation (\ref{quadsys}), (\ref{erer}) that we have proposed for the quad equation (\ref{ge}) is autonomous, but there is the possibility to make a non-autonomous reformulation of the model in which signs of discriminant terms in (\ref{quadsys}) change from quad to quad. 
Such non-autonomous reformulation is more difficult to study, however in the autonomous case the quad equation and its reformulation are not globally equivalent.
They are of course locally equivalent, by which we mean that on a single quad the equation (\ref{ge}) is sufficient for consistency of system (\ref{quadsys}) in $\sigma_1$, $\sigma_2$, allowing the edge variables to be constructed from vertex variables on a single quad.
The simplest setting in which the quad equation and its autonomous reformulation are not equivalent is a square domain involving nine points (four quads).
On this larger domain we can write the system
\begin{equation}
\eqalign{
\cF(u,\wt{u},\wh{u},\th{u},\sigma_1\sigma_2)=0,\\
\cF(u,\ut{u},\wh{u},\ut{\wh{u}},\ut{\sigma}_1\sigma_2)=0,\\
\cF(u,\wt{u},\uh{u},\uh{\wt{u}},\sigma_1\uh{\sigma}_2)=0,\\
\cF(u,\ut{u},\uh{u},\ut{\uh{u}},\ut{\sigma}_1\uh{\sigma}_2)=0.
}
\label{crossys}
\end{equation}
where $\ut{u}=u(n-1,m)$, $\uh{u}=u(n,m-1)$ etc, this system being a consequence of imposing (\ref{quadsys}).
The equation (\ref{ge}) imposed on each quad is {\it not} sufficient for consistency of (\ref{crossys}), specifically elimination of $\sigma_1,\sigma_2,\ut{\sigma}_1$ and $\uh{\sigma}_2$ from (\ref{crossys}) yields a constraint on the nine vertex-variables which is not a consequence of (\ref{ge}).
In particular (\ref{crossys}) can be used to obtain $\th{u}$ uniquely from data $\{\ut{\uh{u}},\ut{u},\uh{u},u,\ut{\wh{u}},\uh{\wt{u}},\wt{u},\wh{u}\}$, whereas if (\ref{ge}) alone were imposed this data would yield two possible values of $\th{u}$.
This example therefore demonstrates that global equivalence of the quad equation and its associated single-valued system is not possible unless a non-autonomous reformulation would be considered.

More generally than the example above, if initial data is specified on verticies along some admissable lattice path, then the autonomous single-valued system (\ref{quadsys}), (\ref{erer}) generically determines $2^{(x-1)}$ solutions where $x$ is the number of edges along the path (corresponding to the number of auxiliary edge variables), whereas using the equation (\ref{ge}) alone the number of solutions determined would be $2^y$ where $y$ is the number of quads in the domain.
Because (on a finite domain) $y\ge x-1$, the autonomous single-valued reformulation of the system can be said to limit the multivaluedness from initial data.

\section{Multidimensional consistency}\label{MDC}
The purpose of this section is to explain a key integrability feature of models (\ref{QQ4})--(\ref{QH2}), namely their {\it multidimensional consistency} \cite{FrankABS,BS,ABS}.
This emerges quite naturally, although it has not been explicitly built into the construction of these equations.

The multidimensional consistency involves not just one equation in isolation, and in fact for the equations listed it involves the whole family of equations obtained by varying parameters $p$ and $q$.
The consistency is between members of this family with different choices of the parameters.
Due to the central role they play it is therefore convenient to explicitly include the dependence on parameters $p$ and $q$ when writing a generic equation from (\ref{QQ4})--(\ref{QH2}):
\begin{equation}
\cQ_{p,q}(u,\wt{u},\wh{u},\th{u})=0.\label{pqs}
\end{equation}

\begin{figure}[t]
\begin{center}
\begin{picture}(200,140)
\input{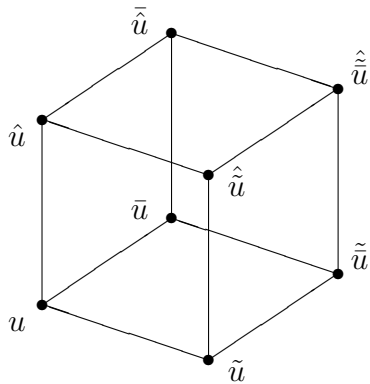}
\end{picture}
\end{center}
\caption{\label{cubepic}Variables assigned to the vertices of a cube. In the case of the multi-quadratic quad equations (\ref{QQ4})--(\ref{QH2}) the system (\ref{cubesys}) determines four possible values of $\thb{u}$ from initial data $u$, $\wt{u}$, $\wh{u}$ and $\wb{u}$.}
\end{figure}

The key properties (involving the parameter dependence) of the generic defining polynomial are first the symmetry
\begin{equation} 
\cQ_{p,q}(u,\wt{u},\wh{u},\th{u}) = \cQ_{q,p}(u,\wh{u},\wt{u},\th{u}), \label{covar}
\end{equation}
and second the consistency of the system
\begin{equation}
\eqalign{
\cQ_{p,q}(u,\wt{u},\wh{u},\th{u})=0, \quad& \cQ_{p,q}(\wb{u},\bt{u},\hb{u},\thb{u})=0,\\
\cQ_{q,r}(u,\wh{u},\wb{u},\hb{u})=0, \quad& \cQ_{q,r}(\wt{u},\th{u},\bt{u},\thb{u})=0,\\
\cQ_{r,p}(u,\wb{u},\wt{u},\bt{u})=0, \quad& \cQ_{r,p}(\wh{u},\hb{u},\th{u},\thb{u})=0,
}\label{cubesys}
\end{equation}
for any choice of the parameters $p$, $q$ and $r$, which is usually visualised by assigning variables to vertices of a cube as in Figure \ref{cubepic}, and equations to faces.
By consistency we mean that for generic initial data $\{u,\wt{u},\wh{u},\wb{u}\}$ the system (\ref{cubesys}) has at least one solution for the remaining variables $\{\th{u},\hb{u},\bt{u},\thb{u}\}$.
Directly by polynomial manipulation it can be confirmed that for each quad-equation (\ref{QQ4})--(\ref{QH2}) the system (\ref{cubesys}) is consistent in this sense, and in fact determines four possible solutions $\{\th{u},\hb{u},\bt{u},\thb{u}\}$.

Alternatively, and more explicitly, the consistency property can be verified by reformulating the equation as a single-valued system as described in Section \ref{SVS}.
For instance in the case of the primary model (\ref{QQ4}) a direct calculation yields the following expression that determines $\thb{u}$ in terms of the initial data on the cube:
\begin{equation}
\fl
\eqalign{
u(c^2-c^{-2})[p(q-r)^2(\wh{u}\wb{u}-\wt{u}\thb{u})+q(r-p)^2(\wb{u}\wt{u}-\wh{u}\thb{u})+r(p-q)^2(\wt{u}\wh{u}-\wb{u}\thb{u})]\\
-(q-r)(r-p)(\wb{u}-\thb{u})[(c^{-1}-cpq)(u^2+\wt{u}\wh{u})+(c^{-1}pq-c)(1+u^2\wt{u}\wh{u})+2\sigma_1\sigma_2]\\
-(r-p)(p-q)(\wt{u}-\thb{u})[(c^{-1}-cqr)(u^2+\wh{u}\wb{u})+(c^{-1}qr-c)(1+u^2\wh{u}\wb{u})+2\sigma_2\sigma_3]\\
-(p-q)(q-r)(\wh{u}-\thb{u})[(c^{-1}-crp)(u^2+\wb{u}\wt{u})+(c^{-1}rp-c)(1+u^2\wb{u}\wt{u})+2\sigma_3\sigma_1]=0,\\
}\label{cubesol}
\end{equation}
where $\sigma_1$, $\sigma_2$ and $\sigma_3$ are determined from the initial data through equations (\ref{dvdef}) and
\begin{equation}
\fl \sigma_3^2 = \frac{c}{2}\left(1+u^2\wb{u}^2+\wb{u}^2r^2+r^2u^2\right)-\frac{1}{2c}\left(u^2\wb{u}^2r^2+u^2+\wb{u}^2+r^2\right)-\left(c^2-\frac{1}{c^2}\right)u\wb{u}r.\label{s3}
\end{equation}
Although the transformation group $(\sigma_1,\sigma_2,\sigma_3)\mapsto(\pm\sigma_1,\pm\sigma_2,\pm\sigma_3)$ leaving (\ref{dvdef}), (\ref{s3}) invariant contains eight elements, the initial data $\{u,\wt{u},\wh{u},\wb{u}\}$ leads to only four distinct values of $\thb{u}$ due to the symmetry $(\sigma_1,\sigma_2,\sigma_3)\mapsto(-\sigma_1,-\sigma_2,-\sigma_3)$ of (\ref{cubesol}).

It is important to note that the consistency property of the associated single-valued system can be broken by reversing sign of all discriminant terms in (\ref{quadsys}).
This subtlety reflects the fact that polynomial system (\ref{cubesys}) determines only four possible values of $\thb{u}$ from the initial data, and not eight as might be expected.
Other polynomials given in Table \ref{svms} are also chosen with the consistency property in mind, so that in particular the sign chosen for the discriminant terms is important.

The consistency property of models (\ref{QQ4})--(\ref{QH2}) means that their more general setting is for a dependent variable $u$ defined on $\mathbb{Z}^d$, where $d>1$ is the dimension, determined by the system
\begin{equation}
\cQ_{p_i,p_j}(u,\oT_iu,\oT_ju,\oT_i\oT_ju)=0, \quad i,j\in\{1\ldots d \}.\label{mds}
\end{equation}
Here $p_1,\ldots,p_d$ are some set of parameters, while $\oT_1,\ldots,\oT_d$ are shift operators in each dimension.
Probably the simplest Cauchy problem for (\ref{mds}) is to specify initial data on coordinate axes, but actually this multidimensional setting is the departure point for an extremely rich variety of initial value problems and lattice configurations \cite{BS,AdVeQ,AS}. 
This also aligns with the notion of solvability for models in statistical mechanics \cite{baxter}.
From a slightly different perspective the consistency property yields a great deal of control over the solution structure of these models by providing immediately a natural auto-B\"acklund transformation.
This allows for example to construct exact solutions as periodic B\"acklund chains (the discrete analogue of finite-gap solutions), as well as soliton-type solutions from B\"acklund iteration.

We remark that on the domain of a hypercube, $\{0,1\}^d$, associated edge variables can be obtained (as described in Section \ref{SVS}) for {\it any} solution of (\ref{mds}), this domain is therefore singled out as one on which the quad equation and its autonomous single-valued reformulation are globally equivalent.

\section{Transformations to ABS equations}\label{transformations}
All equations listed by ABS in \cite{ABS}, with the exception of Q4, are mutually related via B\"acklund or Miura type transformations \cite{NC,James,NAH-Sol,James2}.
Table \ref{bts} here lists similar transformations connecting all except the primary multi-quadratic model Q4$^*$, specifically equations (\ref{QQ3})--(\ref{QH2}), back to multi-affine equations from the ABS list.
Before describing these transformations in more detail we first recall the ABS list in (\ref{Q4})--(\ref{H1}) below.
\subsection{The ABS list}\label{ABSlist}
Q4:
\begin{equation}
\eqalign{
\fl\quad \sn(\alpha)(v\wt{v}+\wh{v}\wh{\wt{v}})-\sn(\beta)(v\wh{v}+\wt{v}\wh{\wt{v}})\\
-\sn(\alpha-\beta)[\wt{v}\wh{v}+v\wh{\wt{v}}-k\sn(\alpha)\sn(\beta)(1+v\wt{v}\wh{v}\wh{\wt{v}})]=0,
}
\label{Q4}
\end{equation}
Q3:
\begin{equation}
\eqalign{
\fl\quad(\alpha-1/\alpha)(v\wt{v}+\wh{v}\th{v})-(\beta-1/\beta)(v\wh{v}+\wt{v}\th{v})\\ 
-(\alpha/\beta-\beta/\alpha)[\wt{v}\wh{v}+v\th{v}+\delta^2(\alpha-1/\alpha)(\beta-1/\beta)/4]=0,
}\label{Q3}
\end{equation}
Q2: 
\begin{equation}
\eqalign{
\fl\quad \alpha(v-\wh{v})(\wt{v}-\th{v})-\beta(v-\wt{v})(\wh{v}-\th{v})\\ 
+\alpha\beta(\alpha-\beta)(v+\wt{v}+\wh{v}+\th{v}-\alpha^2+\alpha\beta-\beta^2)=0,
}\label{Q2}
\end{equation}
Q1:
\begin{equation}
\alpha(v-\wh{v})(\wt{v}-\th{v})-\beta(v-\wt{v})(\wh{v}-\th{v})+\delta^2 \alpha\beta(\alpha-\beta)=0,
\label{Q1}
\end{equation}
A2:
\begin{equation} 
\fl\quad (\alpha-1/\alpha)(v\wh{v}+\wt{v}\th{v})-(\beta-1/\beta)(v\wt{v}+\wh{v}\th{v}) - (\alpha/\beta-\beta/\alpha)(1+v\wt{v}\wh{v}\th{v})=0,
\label{A2}
\end{equation}
A1:
\begin{equation}
\alpha(v+\wh{v})(\wt{v}+\th{v})-\beta(v+\wt{v})(\wh{v}+\th{v})- \delta^2\alpha\beta(\alpha-\beta)=0,
\label{A1}
\end{equation}
H3:
\begin{equation}
\alpha(v\wt{v}+\wh{v}\th{v})-\beta(v\wh{v}+\wt{v}\th{v})+\delta(\alpha^2-\beta^2)=0,
\label{H3}
\end{equation}
H2:
\begin{equation}
(v-\th{v})(\wt{v}-\wh{v})-(\alpha-\beta)(v+\wt{v}+\wh{v}+\th{v}+\alpha+\beta)=0,
\label{H2}
\end{equation}
H1:
\begin{equation}
(v-\th{v})(\wt{v}-\wh{v})-\alpha+\beta=0.
\label{H1}
\end{equation}

In (\ref{Q4})--(\ref{H1}) we have reproduced the ABS list.
Where it appears $\delta\in\{0,1\}$ and for Q4 (\ref{Q4}) $k\in \mathbb{C}\setminus\{0,1,-1\}$ is the modulus of the Jacobi elliptic function $\sn$.
The parametrisations coincide with the ones given by ABS in \cite{ABS} except for the case of Q4, this canonical form (the Jacobi form) was obtained by Hietarinta in \cite{Hie}.
Each ABS equation is defined by a polynomial $\cQ=\cQ(v,\wt{v},\wh{v},\th{v})$ that is degree one in four variables.
It was shown in \cite{ABS} that this polynomial is characterised in terms of biquadratics through a generalised discriminant formula as follows
\begin{equation}
\fl\eqalign{
(\partial_{\wh{v}}\cQ)(\partial_{\th{v}}\cQ)-(\partial_{\wh{v}}\partial_{\th{v}}\cQ)\cQ \propto H_1(v,\wt{v}), \quad (\partial_v\cQ)(\partial_{\wt{v}}\cQ)-(\partial_v\partial_{\wt{v}}\cQ)\cQ  \propto H_1(\wh{v},\th{v}),\\
(\partial_{\wt{v}}\cQ)(\partial_{\th{v}}\cQ)-(\partial_{\wt{v}}\partial_{\th{v}}\cQ)\cQ \propto H_2(v,\wh{v}), \quad (\partial_v\cQ)(\partial_{\wh{v}}\cQ)-(\partial_v\partial_{\wh{v}}\cQ)\cQ  \propto H_2(\wt{v},\th{v}).
}\label{gendisc}
\end{equation}
The polynomials $H_1$ and $H_2$ appearing in (\ref{gendisc}) coincide with those in (\ref{disc}) up to an overall constant, provided we make the following associations between parameters $\alpha$ and $\beta$ appearing in the multi-affine equations (\ref{Q4})--(\ref{H2}) and parameters $p$ and $q$ appearing in their multi-quadratic counterparts (\ref{QQ4})--(\ref{QH2}),
\begin{equation}
\begin{array}{rlll}
Q4:& p=\sqrt{k}\sn(\alpha+K),& q=\sqrt{k}\sn(\beta+K),& c=\sqrt{k},\\
Q3,A2:& p=(\alpha+1/\alpha)/2,&q=(\beta+1/\beta)/2,\\
Q2:& p=\alpha^2,&q=\beta^2,\\
Q1,A1:& p=\delta^2\alpha^2,&q=\delta^2\beta^2,\\
H3:& p=1/\alpha,&q=1/\beta,& \delta^2\rightarrow \delta,\\
H2:& p=\alpha,&q=\beta,
\end{array}
\label{pa}
\end{equation}
where in the case of Q4 the parameter $K$ is standard notation for the quarter period of the $\sn$ function (with modulus $k$) satisfying $\sn(K)=1$, $\sn'(K)=0$.

We remark that for the multi-affine equations H1 (the lattice potential KdV equation \cite{we,NC}), Q1$_{\delta=0}$ (the lattice Schwarzian KdV equation \cite{NC}) and A1$_{\delta=0}$ the associated biquadratic is the square of a polynomial which is degree one in each variable, and as mentioned before their multi-quadratic counterparts factorise into products of multi-affine polynomials which we have excluded as degenerate cases here.

\subsection{An example transformation}\label{BTex}
\begin{table}[t]
\begin{tabular}{lll}
Eq. in $u$ & B\"acklund transformation & Eq. in $v$ \\
\hline
Q3$^*_{p=2\alpha^2-1}$ & $\alpha[(\wt{u}+\delta)\wt{v}^2+(u+\delta)v^2]=(\wt{u}+u+2\delta\alpha^2)\wt{v}v$ & Q3$_{\delta=0}$ \\
Q2$^*_{p=\alpha}$ & $(\wt{v}-v)(\wt{u}\wt{v}-uv)+\alpha \wt{v}v=0$ & Q1$_{\delta=0}$\\
A2$^*_{p=2(\alpha^2+1)/(\alpha^2-1)}$ &$\alpha (\wt{v}^2v^2+1)=\wt{v}v[\wt{u}u(\alpha^2-1)+(\alpha^2+1)]$ & A2\\
H3$^*_{p=4/\alpha^2}$ & $\wt{u}u=\wt{v}v(\wt{v}v+\delta\alpha)$ & H3\\
H3$^*_{p= 4/\alpha}$ & $4\alpha\wt{u}u=(\wt{v}+v)^2-\delta^2\alpha^2$ & A1 \\
A1$^*_{p=\alpha^2}$ & $2 \wt{v} v(\wt{u}+u)=\alpha(\wt{v}^2 v^2+1)$ & H3$_{\delta=0}$\\
A1$^*_{p=\alpha}$  & $2 (\wt{v}+v)(\wt{u}+u)= (\wt{v}+v)^2+\alpha$ & A1$_{\delta=0}$\\
H2$^*_{p=-\alpha}$ & $\wt{u}+u-\alpha=(\wt{v}+v)^2$ & H1 \\
\hline
A2$^*_{p=2\alpha^2-1}$ & $\alpha(\wt{u}u\wt{v}^2v^2+1)=(\wt{u}u+1)\wt{v}v$ & A2\\
H3$^*_{p=4/\alpha}$ & $4\alpha \wt{u}u=(\wt{v}-v)^2-\delta^2\alpha^2$ & Q1 \\
Q1$^*_{p=\alpha^2}$ & $2\wt{v}v(\wt{u}-u)=\alpha( \wt{v}^2 v^2+1)$ & H3$_{\delta=0}$\\
Q1$^*_{p=\alpha}$ & $ 2(\wt{v}-v)(\wt{u}-u)=(\wt{v}-v)^2+\alpha$ & Q1$_{\delta=0}$\\
\hline
\end{tabular}
\caption{B\"acklund transformations connecting equations from the multi-quadratic class to the multi-affine (ABS) class. 
Details of implementing the transformations are explained in Section \ref{BTex}.
Transformations are given up to composition with point symmetries of the equations in $u$ and $v$.
For completeness we include in the second part of the table some transformations that can be obtained from those in the first part by composition with (non-autonomous) point transformations.
The transformation connecting H2$^*$ to H1 was given originally in \cite{AdVeQ} up to composition with point symmetries.
Methods to obtain the transformations are described in the text, first by non-symmetric degeneration of the auto-B\"acklund transformation in Section \ref{nsd}, second by taking advantage of a natural connection with Yang-Baxter maps in Section \ref{idolons}, and third in Section \ref{quadlin} by exploiting the fact that the transformations, like the equations themselves, can be characterised by discriminant properties of the defining polynomial.
}\label{bts}
\end{table}
To explain precisely the meaning of the entries in Table \ref{bts} we give here an explicit example, specifically the second entry of the table.
Consider the coupled system of equations
\begin{equation}
(v-\wt{v})(uv-\wt{u}\wt{v})+\alpha v \wt{v} = 0, \quad (v-\wh{v})(uv-\wh{u}\wh{v})+\beta v\wh{v} = 0,\label{bt}
\end{equation}
which involve two functions $u=u(n,m)$ and $v=v(n,m)$, where as usual $n,m\in\mathbb{Z}$, $\wt{u}=u(n+1,m)$ and $\wh{u}=u(n,m+1)$, etc.

For a fixed function $v=v(n,m)$ the equations (\ref{bt}) are coupled discrete Riccati equations for $u$, which are compatible if $v$ satisfies Q1$_{\delta=0}$.
Due this choice of $v$ the function $u=u(n,m)$ that emerges as the solution of (\ref{bt}) then satisfies equation Q2$^*$, that is (\ref{QQ2}), with parameter associations $p=\alpha$ and $q=\beta$ (notice that this parameter association is different from the one in (\ref{pa})).
For brevity the second equation in (\ref{bt}) and the parameter association between $q$ and $\beta$ are omitted from Table \ref{bts} because they can be inferred from the first equation in (\ref{bt}) and the association between $p$ and $\alpha$ that have been listed.

The transformation defined by (\ref{bt}) from $v$ to $u$ is therefore a quite standard discrete-Riccati type of B\"acklund transformation, see for instance \cite{NC,BS,James}.
However the system (\ref{bt}) also defines an inverse transformation from $u$ to $v$ which is less standard because the equations for $v$ are not of Riccati type.
Similar to verifying the consistency property of the multi-quadratic models (\ref{QQ4})--(\ref{QH2}) (cf. Section \ref{MDC}), it is convenient to handle this system by introducing auxiliary variables.
Specifically for the example (\ref{bt}) the auxiliary variables enter through the edge relations
\begin{equation}
\eqalign{
\sigma_1^2 = \frac{1}{4}\left(u^2+\wt{u}^2+\alpha^2\right)-\frac{1}{2}\left(u\wt{u}+\wt{u}\alpha+\alpha u\right),\\
\sigma_2^2 = \frac{1}{4}\left(u^2+\wh{u}^2+\beta^2\right)-\frac{1}{2}\left(u\wh{u}+\wh{u}\beta+\beta u\right),
}
\end{equation}
so in fact they are the Q2$^*$ auxiliary variables.
They allow (\ref{bt}) to be re-written as
\begin{equation}
2\wt{u}\wt{v}=(u+\wt{u}-\alpha-2\sigma_1)v, \quad 2\wh{u}\wh{v}=(u+\wh{u}-\beta-2\sigma_2)v, \label{bt2}
\end{equation}
which have degree-one polynomial dependence in $v$, $\wt{v}$ and $\wh{v}$.
Implementation of the transformation from $u$ to $v$ therefore requires a solution $(u,\sigma_1,\sigma_2)$ of the single-valued reformulation of Q2$^*$ (cf. Table \ref{svms}).
From such solution the system (\ref{bt2}) determines the function $v$ satisfying Q1$_{\delta=0}$.

Similar to the example we have focused on in this section, all transformations listed in Table \ref{bts} are standard Riccati-type systems for $u$.
The inverse transformation to obtain $v$ is always more involved, reducing to a Riccati type system for $v$ only after the introduction of auxiliary variables associated with the solution of the multi-quadratic quad equation for $u$.

\subsection{Obtaining the transformations}\label{nsd}
It is rare to have a systematic method to obtain non-local transformations of B\"acklund or Miura type between a given pair of equations.
The transformations obtained here do however fit into a general framework, specifically they are defined in terms of polynomials which are characterised by their discriminants, a framework which is therefore consistent with the main theme of this paper. 
We exploit this general point of view in Section \ref{quadlin} however, we instead use more direct methods to obtain most of the transformations listed in Table \ref{bts}.
A method involving {\it non-symmetric degeneration} of the auto-B\"acklund transformation is explained in this section, transformations obtained through a connection with Yang-Baxter maps will be explained in detail in the following section (Section \ref{idolons}).

There is a method to obtain B\"acklund transformations between distinct quad equations that was developed in \cite{James}, it is constructive insofar as it takes as a starting point the already obtained equations.
It relies on the natural auto-B\"acklund transformation, which for the equations in question is inherent from the multidimensional consistency, and exploits this in conjunction with the hierarchical relationships between equations. 
Specifically we seek to connect an equation to its degenerate counterpart by making a non-symmetric degeneration of the auto-B\"acklund transformation.

The principal example where the non-symmetric degeneration technique has been used is to obtain the second transformation in Table \ref{bts} (which was also the example considered in Section \ref{BTex}).
We exploit the fact that the substitution
\begin{equation}
u=\frac{v}{\epsilon(1+v)}
\end{equation}
into equation Q2$^*$ yields equation Q1$_{\delta=0}$ to first order as $\epsilon \longrightarrow 0$.
The natural auto-B\"acklund transformation for Q2$^*$ (\ref{QQ2}) is as follows
\begin{equation}
\cQ_{p,r}(u,\wt{u},v,\wt{v})=0, \quad \cQ_{q,r}(u,\wh{u},v,\wh{v})=0,
\end{equation}
where $\cQ_{p,q}$ is the defining polynomial of this equation, and the transformation connects a solution $u=u(n,m)$ to another solution of the same equation $v=v(n,m)$.
Applying the degeneration procedure to solution $v$ and judiciously choosing the B\"acklund parameter we are led to write
\begin{equation}
\fl \cQ_{p,r}(u,\wt{u},rv/(1+v),r\wt{v}/(1+\wt{v}))=0, \quad \cQ_{q,r}(u,\wh{u},rv/(1+v),r\wh{v}/(1+\wh{v}))=0,
\end{equation}
which at leading order as $r=1/\epsilon\longrightarrow \infty$ yields exactly the desired non-auto B\"acklund transformation (\ref{bt}).

\subsection{From Yang-Baxter maps to B\"acklund transformations}
\label{idolons}
The Yang-Baxter maps given in \cite{ABSf}, when suitably interpreted as a system of equations for functions defined on edges of the lattice, are naturally connected with multi-affine quad equations from the ABS list \cite{ABS}, in particular a potential function for the edge variables is governed by a quad equation. 
A systematic method to obtain the Yang-Baxter system on edge variables starting from the multi-affine equations from the ABS list was given in \cite{Tasos}.
Developments reported in \cite{KaNie,KaNie3} go further, but in the opposite direction, showing that also non-multi-affine multidimensionally consistent quad equations can emerge as potential for the Yang-Baxter systems.
And furthermore, that two different quad equations emerging in this way from the same Yang-Baxter system, as we shall see in this section, can immediately be connected through a B\"acklund-type transformation.

Of particular relevance here are the models that have been considered in \cite{KaNie5} alongside the present work.
This section is devoted to recalling the relevant models from \cite{ABSf} and giving details of the theory developed in \cite{KaNie,KaNie3,KaNie5} which, combined with the methods developed in this paper, can be used to obtain many of the transformations in Table \ref{bts}.

The starting point is a system of equations for two variables, say $s$ and $t$, assigned to the edges of $\mathbb{Z}^2$ oriented in the $n$ and $m$ directions respectively (similar to $\sigma_1$ and $\sigma_2$ introduced in Section \ref{SVS}, cf. Figure \ref{quadpic}).
The particular systems relevant here are as follows.

{\noindent $F_I$:}
\begin{equation}
\label{A2I}
\eqalign{
\wh{s} = t\frac{\alpha (1-{\beta}^2) s - \beta (1-{\alpha}^2) t-{\alpha}^2+{\beta}^2}{\beta (1-{\alpha}^2) s - \alpha (1-{\beta}^2) t+({\alpha}^2-{\beta}^2)st}, \\
\wt{t} = s\frac{\beta (1-{\alpha}^2) t - \alpha (1-{\beta}^2) s-{\beta}^2+{\alpha}^2}{\alpha (1-{\beta}^2) t - \beta (1-{\alpha}^2) s+({\beta}^2-{\alpha}^2)st},
}
\end{equation}
$F_{II}$:
\begin{equation}
\label{II}
\eqalign{
\wh{s}= t\frac{\alpha s-\beta t-\delta(\alpha^2-\beta^2)}{\beta s-\alpha t},\\
\wt{t}= s\frac{\alpha s-\beta t-\delta(\alpha^2-\beta^2)}{\beta s-\alpha t},
}
\end{equation}
$F_{III}$:
\begin{equation}
\label{III}
\eqalign{
\wh{s}= t \frac{\alpha s-\beta t}{\beta s-\alpha t},\\
\wt{t}= s \frac{\alpha s-\beta t}{\beta s-\alpha t},
}
\end{equation}
$F_{V}$:
\begin{equation}
\label{V}
\eqalign{
\wh{s}=t+\frac{\alpha-\beta}{s-t},\\
\wt{t}=s+\frac{\alpha-\beta}{s-t},
}
\end{equation}
which are nothing but the quadrirational Yang-Baxter maps presented in \cite{ABSf} suitably interpreted as an equation on the lattice (we have omitted model $F_{IV}$ because it is not used here).
The relevant feature of systems (\ref{A2I})--(\ref{V}) is that they admit a three-parameter family of potentials, we denote the potential here by $f$.
The potentials corresponding to these models, which are derived in \cite{KaNie5}, are as follows:

\begin{equation}
\label{A2p}
\fl (F_{I})\quad
\begin{array}{l}
\wt{f}+f = A\ln(s)+B\ln(s-\alpha)+C\ln(\alpha s-1)+\frac{1}{2}(B+C)\ln({\alpha}^2-1),\\
\wh{f}+f = A\ln(t)+B\ln(t-\beta)+C\ln(\beta t-1)+\frac{1}{2}(B+C)\ln({\beta}^2-1),
\end{array}
\end{equation}
\begin{equation}
\fl (F_{II})\quad
\begin{array}{l}
\wt{f}+f=A\alpha (2s-\delta\alpha)+B \ln(s) +C \ln (s-\delta\alpha),\\
\wh{f}+f=A\beta (2t-\delta\beta)+B \ln(t) +C \ln (t-\delta\beta),
\end{array}
\end{equation}
\begin{equation}
\fl (F_{III})\quad
\begin{array}{l}
\wt{f}+f=A\ln(s)+\alpha B s+\displaystyle {C}/{s},\\
\wh{f}+f=A\ln(t)+\beta B t+\displaystyle {C}/{t},
\end{array}
\end{equation}
\begin{equation}
\fl (F_{V})\quad
\begin{array}{l}
\wt{f}+f=As+B(s^2+\alpha)+C(s^3+3 \alpha s),\\
\wh{f}+f=At+B(t^2+\beta)+C(t^3+3 \beta t),
\end{array}
\label{Vp}
\end{equation}
The parameters $A$, $B$ and $C$ may be chosen freely in each case.

For the purpose of illustration we focus on the system (\ref{A2I}).
In particular writing $f=\ln(v)$ for the potential corresponding to the choice of the parameters $A=1$, $B=C=0$ in (\ref{A2p}) we find
\begin{equation}
\label{vv}
\wt{v}v = s, \quad \wh{v}v = t,
\end{equation}
whilst writing $f=\ln(u)$ for the potential corresponding to the choice of the parameters $A=-1$, $B=C=1$, (\ref{A2p}) becomes
\begin{equation}
\label{uu}
\wt{u}u =\frac{(s-\alpha)(\alpha s-1)}{s({\alpha}^2-1)}, \quad
\wh{u}u =\frac{(t-\beta)(\beta t-1)}{t({\beta}^2-1)}.
\end{equation}
Eliminating $s$ and $t$ from (\ref{A2I}) by means of (\ref{vv}) we find that $v$ satisfies equation A2 (\ref{A2}), whilst eliminating $s$ and $t$ from (\ref{A2I}) using (\ref{uu}) we find that variable $u$ satisfies equation A2$^*$ (\ref{QA2}) with the parameter associations
\[p =2\frac{\alpha^2+1}{\alpha^2-1},\quad q =2\frac{\beta^2+1}{\beta^2-1}.\]
Equations governing the potential $f$ for different choices of the parameters $A$, $B$ and $C$ are referred to in \cite{KaNie} as {\em idolons} of the underlying system governing $s$ and $t$, in particular we have shown that equations A2 and A2$^*$ are idolons of (\ref{A2I}).

A B\"acklund transformation can be obtained by composing these relations, that is by eliminating $s$ and $t$ from formulas (\ref{vv}) and (\ref{uu}),
\begin{equation}
\label{BTI}
\wt{u}u =\frac{(\wt{v}v-\alpha)(\alpha \wt{v}v-1)}{\wt{v}v({\alpha}^2-1)}, \quad
\wh{u}u =\frac{(\wh{v}v-\beta)(\beta \wh{v}v-1)}{\wh{v}v({\beta}^2-1)}.
\end{equation}
This is the B\"acklund transformation between $A2$ and $A2^*$ that appears in the first part of Table \ref{bts}.
The second transformation between these models can be obtained similarly after observing that the potential corresponding to choice of parameters $A=C=-1$, $B=1$ is also governed by model $A2^*$. 

Table \ref{mtab} contains the data required to construct, by the same procedure, all but the first two transformations in the first part of Table \ref{bts}.
\begin{table}
\begin{tabular}{llll}
System & Parameters ($A$,$B$,$C$) & Potential & Equation\\
\hline
$F_{I}$ & (-1,1,1) & $f=\ln(u)$ & A2$^*_{p=2(\alpha^2+1)/(\alpha^2-1)}$\\
        & (-1,1,-1) & $f=\ln(u)$ & A2$^*_{p=2\alpha^2-1}$\\
        & (1,0,0) & $f=\ln(v)$ & A2\\
\hline
$F_{II}$ & (0,1,1) & $f=\ln(u)$ & H3$^*_{p=4/\alpha^2}$\\
         & (0,0,1) & $f=\ln(v)$ & H3\\
         & (1,0,0) & $f=v$ & A1$_{\alpha\rightarrow \alpha^2}$\\
\hline
$F_{III}$ & (0,\textonehalf,\textonehalf) & $f=u$ & A1$^*_{p=\alpha^2}$\\
          & (1,0,0) & $f=\ln(v)$ & H3$_{\delta=0}$\\
          & (0,1,0) & $f=v$ & A1$_{\delta=0,\alpha\rightarrow \alpha^2}$\\
\hline
$F_{V}$ & (0,1,0) & $f=u$ & H2$^*_{p=-\alpha}$\\
        & (1,0,0) & $f=v$ & H1\\
\end{tabular}
\caption{
Equations governing different potentials of the systems (\ref{A2I})--(\ref{V}).
The potentials are introduced through equations (\ref{A2p})--(\ref{Vp}) with the indicated choice of parameters $(A,B,C)$.
It is explained in the text how to construct a transformation between equations (idolons) that govern different potentials of the same system.
Parameter associations in the listed equations are given connecting $\alpha$ and $p$ or transforming $\alpha$, similar associations connecting $\beta$ and $q$ or transforming $\beta$ are implicit, they are omitted from the table for brevity.
}
\label{mtab}
\end{table}

\subsection{Discriminant properties of the transformations}
\label{quadlin}
All transformations listed in Table \ref{bts} are in the same class, they involve equations on lattice edges and the defining polynomial $\cB=\cB(u,\wt{u},v,\wt{v})$ is degree-one in each of $u$ and $\wt{u}$, and degree two in each of $v$ and $\wt{v}$.
This defining polynomial also has the following discriminant properties,
\begin{equation}
\eqalign{
(\partial_{u}\cB)(\partial_{\wt{u}}\cB)-(\partial_u\partial_{\wt{u}} \cB)\cB \propto \mu{\wt{\mu}} h(v,\wt{v}),\\
(\partial_{\wt{v}}\cB)^2-2({\partial^2_{\wt{v}}} \cB)\cB \propto \eta^2 h^*(u,\wt{u}),
}\label{ddd}
\end{equation}
where $\mu$ and $\eta$ are polynomials in $v$, and $h$, $h^*$ are the edge biquadratics associated with the two equations connected by the B\"acklund transformation.
Specifically $h$ coincides with $H_1$ associated with the multi-affine equation in $v$ through the generalised discriminant (\ref{gendisc}), while $h^*$ coincides with $H_1$ associated with the multi-quadratic equation in $u$ through the discriminant formula (\ref{disc}).
This discriminant property is an important feature of the transformations that combines asymmetrically the underlying discriminant characterisations of the multi-quadratic models (\ref{QQ3})--(\ref{QH2}) and the multi-affine ABS equations (\ref{Q3})--(\ref{H1}). 
In particular it may be used as a basis for their construction.

One approach to such construction is offered by recognising the biquadratics themselves (\ref{q4})--(\ref{h2}) can take natural discriminant forms, as the primary example we recognise that (\ref{q3}) is proportional to 
\[ (\wt{u}+u+2\delta \alpha^2)^2-4[\alpha(\wt{u}+\delta)][\alpha(u+\delta)], \qquad 2\alpha^2=p+1. \]
It is easily seen that this expression emerges as the discriminant, with respect to $\wt{v}$ or $v$, of the expression
\[ [\alpha(\wt{u}+\delta)]\wt{v}^2 - [\wt{u}+u+2\delta\alpha^2]\wt{v}v + [\alpha(u+\delta)]v^2, \]
which itself coincides with the polynomial defining the first transformation in Table \ref{bts}.
The precise choice of this expression is not unique, and in particular the biquadratic obtained from the generalised discriminant in $u$ and $\wt{u}$ emerges a posteriori after choosing the expression on the basis that the required discriminant property in $v$ and $\wt{v}$ is obtained.

We remark that the classification of polynomials $\cB$ with the discriminant properties (\ref{ddd}) is interesting from the point of view of exhausting transformations in the same class as those listed in Table \ref{bts} (such task is similar to the one solved for multi-affine quad equations in \cite{boll}).

\section{Discussion}\label{discussion}
The discriminant factorisation property, as laid out in Section \ref{dfh}, allows reformulation of the multi-quadratic quad equation as a system that defines single-valued evolution from initial data.
Our main result is the list of models constructed on the basis of this hypothesis in Section \ref{list}, it contains all previously known integrable multi-quadratic quad equations as well as a substantial number of new models that exhibit the same integrability features.
The nature of the relationship between the discriminant factorisation property and the integrability is therefore an important question.
Here we have made additional assumptions beyond the discriminant factorisation hypothesis, therefore although our investigations are suggestive, they leave open the problem of determining whether this property is sufficient for integrability in this class of models.

Beyond elliptic (and hyperelliptic) functions, degree-two equations studied systematically from the point of view of integrability seem to be relatively rare.
An isolated precedent exists within the framework of Painlev\'e analysis for ordinary differential equations.
A list of degree-two counterparts of the Painlev\'e equations was obtained by Chazy \cite{chsy}, loosely speaking these are second-order ordinary differential equations with the Painlev\'e property that are quadratic in the second derivative term.
The difficult step of rigorously classifying this class of equations was made by Cosgrove and Scoufis \cite{CoSc}.
In that setting one of the principal features is that the higher degree models do not define new transcendents (see \cite{Cosg}), in particular they are solvable in terms of the degree-one Painlev\'e equations.
Here the primary model, namely the equation that we identify as a multi-quadratic counterpart of Q4, is the only new model we have not yet connected back to an equation from the multi-affine class.
The parallels between integrable quad-equations and the Painlev\'e-type equations suggest that a B\"acklund-type transformation establishing such connection should exist.
Verifying or falsifying this is therefore an important open problem.

\ack
J.A is supported by the Australian Research Council, Discovery Grant DP 110104151.
M.N is grateful for financial support from Australian Research Council Discovery Grant DP 0985615, which enabled his stay at the University of Sydney, during which the present surveys started.

\section*{References}
\bibliographystyle{unsrt}
\bibliography{biblio-ref}
\end{document}